\documentclass[12pt]{raa}            % referee version: for submission
\usepackage{graphicx,times}
\usepackage{natbib}

\providecommand{\bm}[1]{\mbox{\boldmath$#1$\unboldmath}}

\newcommand{\Chandra}{{\it Chandra}}

\newcommand{\ASCA}{{\it ASCA}}

\usepackage{epsfig}

\begin{document}

\title{Metal Enrichment via Ram Pressure Stripping in the IGM of the
  Compact Galaxy Group RGH 80}

\author{Haijuan Cui\inst{}
  \and
  Haiguang Xu\inst{}
  \and 
  Junhua Gu\inst{}
  \and 
  Jingying Wang\inst{}
  \and 
  Liyi Gu\inst{}
  \and 
  Yu Wang\inst{}
  \and
  Zhenzhen Qin\inst{}
}

\institute{Department of Physics, Shanghai Jiao Tong University, 800 Dongchuan Road, Shanghai 200240, China; {\it hgxu@sjtu.edu.cn}
}

\abstract{ By creating and analyzing the two dimensional gas
  temperature and abundance maps of the RGH 80 compact galaxy group
  with the high-quality \Chandra\ data, we detect a high-abundance
  ($\simeq 0.7$ $Z_\odot$) arc, where the metal abundance is
  significantly higher than the surrounding regions by $\simeq 0.3$
  $Z_\odot$. This structure shows tight spatial correlations with the
  member galaxy PGC 046529, as well as with the arm-like feature
  identified on the X-ray image in the previous work of Randall et
  al. (2009).  Since no apparent signature of AGN activity is found
  associated with PGC 046529 in multi-band observations, and the gas
  temperature, metallicity, and mass of the high-abundance arc
  resemble those of the ISM of typical early-type galaxies, we
  conclude that this high-abundance structure is the remnant of the
  ISM of PGC 046529, which was stripped out of the galaxy by ram
  pressure stripping due to the motion of PGC 046529 in RGH 80. This
  novel case shows that ram pressure stripping can work efficiently in
  the metal enrichment process in galaxy groups, as it can in galaxy
  clusters.
   }

\authorrunning{H. Cui et al.}  %author_head in even pages
\titlerunning{Metal Enrichment In RGH 80} % title_head in odd pages
\maketitle

% #####################
\section{Introduction}
% #####################
In galaxy groups and clusters, the inter-galactic medium (IGM) is
enriched with the metals synthesized in the stellar interior, which
were released into the vast inter-galactic space mainly via mergers
(e.g., Durret et al. 2005), galactic wind (e.g., Nath \& Trentham
1997), and AGN activities (e.g., Simionescu et al. 2009). As of today,
the details of how these mechanisms work and how effective they are at
different evolution stages of the IGM are still unclear (e.g., Aguirre
et al. 2001; Hayakawa et al. 2004, 2006). Although AGN feedback has
been regarded as the most popular and effective process to enrich the
IGM, the works of e.g., Hayakawa et al. (2004, 2006) still showed that
ram pressure stripping caused by merger may have played a crucial role
in removing metals out of galaxies.

Very few cases of ram pressure stripping of metal-enriched gas
contained in galaxies have been reported in literature (e.g., Sun et
al. 2007; Hayakawa et al. 2004, 2006), and all these cases were found
in galaxy clusters, where the dynamical environment is
complicated. However, in compact galaxy groups where there exist much
fewer member galaxies as compared with the clusters, the galaxy number
density is actually as high as that of the core region of a rich
cluster, so that strong interactions between member galaxies are often
observed (e.g., Carlberg et al. 1994). This indicates that compact
galaxy groups are ideal sites for us to study the effects of ram
pressure stripping of metal-enriched gas held in the member galaxies.

The compact galaxy group RGH 80 ($z=0.0377$) may be such an ideal
place. In 2009, Randall et al. reported that there exist interactions
between the dominating galaxy PGC 046515 (E, Vorontsov-Velyaminov et
al. 2001; R.A.=13h20m14.8s, Dec=+33d08m37.7s, J2000, Brinkmann et
al. 2000) and the member galaxy PGC 046529 (E, Vorontsov-Velyaminov et
al. 2001; R.A.=13h20m17.8, Dec=+33d08m41.2, J2000, Rines et al. 2003),
and ascribed the origin of an arm-like X-ray feature identified on the
X-ray map to the ram pressure stripping. However, a detailed analysis
of the two dimensional distribution of gas metallicity, which is of
fundamental importance to study the origin of the X-ray arm was not
provided by the authors.

In this work, we calculate and study the two dimensional spatial
distributions of both gas temperature and abundance of the hot IGM of
RGH 80 with the high-quality \Chandra\ data archived at the \Chandra\
X-ray Observatory Science Center that is operated by the Smithsonian
Astrophysical Observatory. We find firm evidence for the IGM of
this group being enriched via the ram pressure stripping caused by the
motion of the member galaxy PGC 046529. Throughout the paper, we quote
errors at 90\% confidence level unless mentioned otherwise. We adopt
the solar abundance standard of Grevesse and Sauval (1998), where the
iron abundance relative to hydrogen is $3.16\times10^{-5}$ in number,
and employ the cosmological parameters $H_0=71h_{71}$ km s$^{-1}$
Mpc$^{-1}$, $\Omega_m=0.27$, $\Omega_\Lambda=0.73$, so that $1^\prime$
corresponds to $44.2h_{71}^{-1}$ kpc.

%#######################################
\section{OBSERVATION AND DATA REDUCTION}
%#######################################

The RGH 80 group was observed with \Chandra\ on November 1, 2005
(ObsID 6941) for a total exposure of 39.1 ks with CCDs 2, 3, 5, 6, and
7 of the Advanced CCD Imaging Spectrometer (ACIS) in operation.  The
center of the group-dominating galaxy PGC 046515 was positioned closed
to the nominal aim point on the back-illuminated S3 chip (CCD 7) with
a small offset of $0.18^{\prime}$.  The events were collected with a
frame time of 3.2 s and telemetried in the VFaint mode, as the focal
plane temperature set to $-120^{\circ}$. In this work we use the
standard \Chandra\ data analysis package CIAO software (version 3.4.0)
and apply the latest CALDB (version 4.1.1) to process the data
extracted from the ACIS-S3 chip.  We keep events with \ASCA\ grades 0,
2, 3, 4 and 6, and remove all the bad pixels, bad columns, columns
adjacent to bad columns and node boundaries. We examine the
lightcurves extracted from the source-free regions on the S3 chip and
detect no strong occasional background flares. Based on above
processings, we obtain a total of about $1.2\times10^{5}$ photons in
the level-2 S3 event list.

\section{X-ray and Optical Images}
In Figures 1{\it a} and 1{\it b} we show the $0.7-7$ keV \Chandra\
ACIS-S3 image that has been corrected for exposure and smoothed with a
Gaussian kernel of $4^{\prime\prime}$, and the corresponding Digitized
Sky Survey (DSS) B-band image of RGH 80, respectively. We find that
the position of the X-ray peak (R.A.=13h20m14.7s, Dec=+33d08m36.3s) is
consistent with the optical center of the galaxy PGC 046515 within
about $2.3^{\prime\prime}$, while another major member galaxy PGC
046529 is located at $0.64^\prime$ east of PGC 046515.  The
distribution of the diffuse X-ray emission is roughly symmetric within
the central $\simeq0.5^{\prime}$ ($22.1h_{71}^{-1}$ kpc). Outside the
central region, we confirm the result of Randall et al. (2009) that
there exists an arm-like X-ray feature in the northeast direction,
which spans from north to east with one of its ends connected to PGC
046529 (Fig. 1{\it a}). To quantitate the significance of this
feature, we calculate the averaged $0.7-7$ keV surface brightness of
the four pie regions, which are defined in Figure 1{\it a}, with the
northeast one covering the arm-like feature. We find that the averaged
surface brightness of the northeast pie region is
$5.1\pm0.2\times10^{-9}$ cts s$^{-1}$ cm$^{-2}$ pixel $^{-2}$, which
is significantly higher than those of other pie regions
($3.2\pm0.1\times10^{-9}$ cts s$^{-1}$ cm$^{-2}$ pixel$^{-2}$,
$3.1\pm0.1\times10^{-9}$ cts s$^{-1}$ cm$^{-2}$ pixel$^{-2}$, and
$3.2\pm0.1\times10^{-9}$ cts s$^{-1}$ cm$^{-2}$ pixel$^{-2}$ for the
southeast, southwest, and northwest pie regions, respectively) at
$90\%$ confidence level. The appearance of this arm-like feature
implies the existence of strong distortion in this region. Using the
group's average temperature $\simeq 1$ keV (Xue et al. 2004), we
calculate the local sound speed to be $\simeq 400$ km s$^{-1}$, which
means that the sound crossing time over the width $\simeq 0.3^\prime$
($13.3h_{71}^{-1}$ kpc) of this feature is $\simeq 30$ Myr, a time
that can be regarded as an approximation to the upper limit of the
feature's lifetime.

%=============
%Fig. 1 Images
%=============
%#############
\begin{figure}[h!!!]
\begin{center}
\includegraphics[width=12.5cm,angle=0]{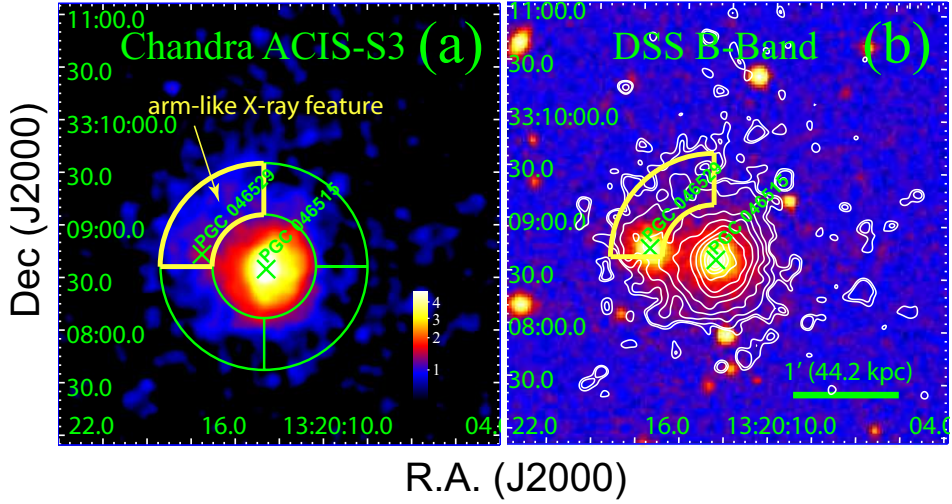}
\end{center}
\caption{(a) $0.7-7$ keV \Chandra\ ACIS-S3 image that has been
  corrected for exposure and smoothed with a Gaussian kernel of
  $4^{\prime\prime}$. The unit of the color bar is $10^{-8}$ cts
  s$^{-1}$ cm$^{-2}$ pixel$^{-2}$. The four pie regions are used to
  extract and compare the surface brightness of the arm-like feature
  with other directions. (b) The DSS B-band optical image of RGH 80
  overlaid with the X-ray intensity contours calculated from (a), with
  the contours ranging from $7\times10^{-9}$ cts s$^{-1}$ cm$^{-2}$
  pixel$^{-2}$ to $6\times10^{-8}$ cts s$^{-1}$ cm$^{-2}$
  pixel$^{-2}$. In both (a) and (b), we mark the positions of PGC
  046515 and PGC 046529 with two crosses and the arm-like X-ray
  feature with the yellow pie region (see also Fig. 1 in Randall et
  al. 2009).}
\end{figure}

\section{2-D Spectral Analysis}

Following the method of Gu et al. (2007), we calculate the two
dimensional temperature and abundance maps of the X-ray gas and their
corresponding $1\sigma$ error maps (Fig. 2). At the first step, we
distribute a set of 1000 sampling knots on the sky plane, with the
surface density distribution of the sampling knots proportional to the
X-ray surface brightness distribution. Next we divide the group with
the CVT cells (Diehl \& Statler 2006) generated around these sampling
knots. We extract the spectrum from a circular region centered on each
sampling knot, which encloses more than 1000 photons in $0.7-7$ keV
(including the background) after all the identified point sources are
excluded. The extracted spectrum is fitted with an absorbed APEC model
(XSPEC v.12.4.0), with the absorption fixed to the Galactic column
density $N_H=1.05\times10^{20}$ cm$^{-2}$ (Dickey \& Lockman 1990),
and the redshift fixed to 0.0377. The \Chandra\ blank field spectra
are used as the background. We assign the obtained temperature and
abundance to the corresponding CVT cell, and repeat this process for
all sampling knots. Finally we obtain the temperature and abundance
maps through adaptively smoothing the tessellated images with a
Gaussian kernel, whose scale is set to be the radii of the CVT
cells. Error maps are created in the similar way.

%=============
%Fig. 2 Images
%=============
%#############
\begin{figure}[h!!!]
\begin{center}
  \includegraphics[width=12cm]{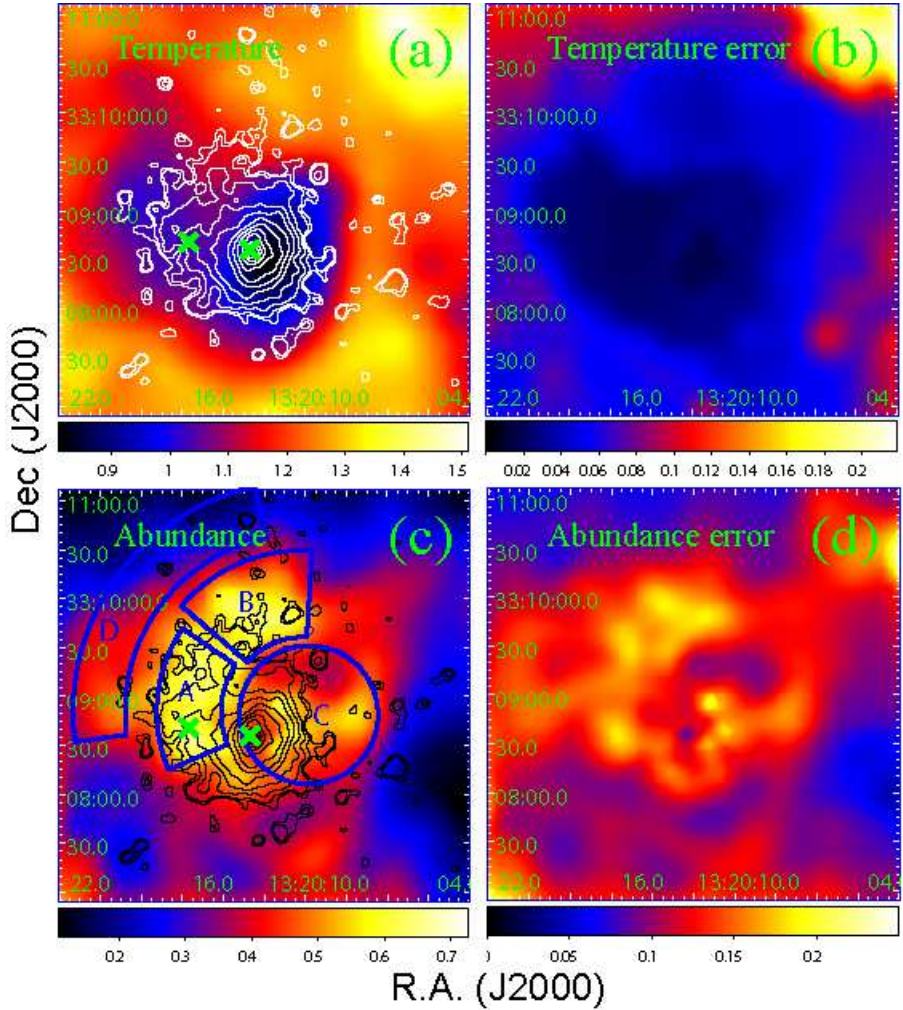}
\end{center}
\caption{(a) The temperature map of RGH 80 overlaid with the X-ray
  contours (Fig. 1{\it b}). (b) The $1\sigma$ error map for the
  temperature distribution. (c) The abundance map of RGH 80 overlaid
  with the X-ray contours (Fig. 1{\it b}). Regions A, B, C, and D are
  used to extract the spectra to confirm the existence of the
  high-abundance arc. (d) The $1\sigma$ error map for the abundance
  distribution.  In both (a) and (c), we mark the position of PGC
  046515 and PGC 046529 with two crosses.}
\end{figure}

We find that both the temperature and abundance distributions appear
to be highly asymmetric. Figure 2{\it a} shows that the center cool
($<1$ keV) region shows an apparent offset from the X-ray peak to the
west, which agrees with the distortions illustrated on the X-ray image
(Fig. 1{\it a}). To the east of the X-ray peak, there is a relatively
low temperature ($\simeq 1$ keV) region, which encloses the member
galaxy PGC 046529. On the abundance map, we identify an impressively
high-abundance arc that spans from north to east, with a radius of
$\simeq 1.24^\prime$ ($54.9h_{71}^{-1}$ kpc) and a width of
$\simeq0.8^\prime$ ($35.4h_{71}^{-1}$ kpc). The abundances in and
around the arc are $\simeq0.7\pm0.16$ $Z_\odot$ and $0.4\pm 0.12$
$Z_\odot$ (68\% confidence level), respectively.

\begin{figure}[h!!!]
\begin{center}
\includegraphics[width=12.5cm,angle=0]{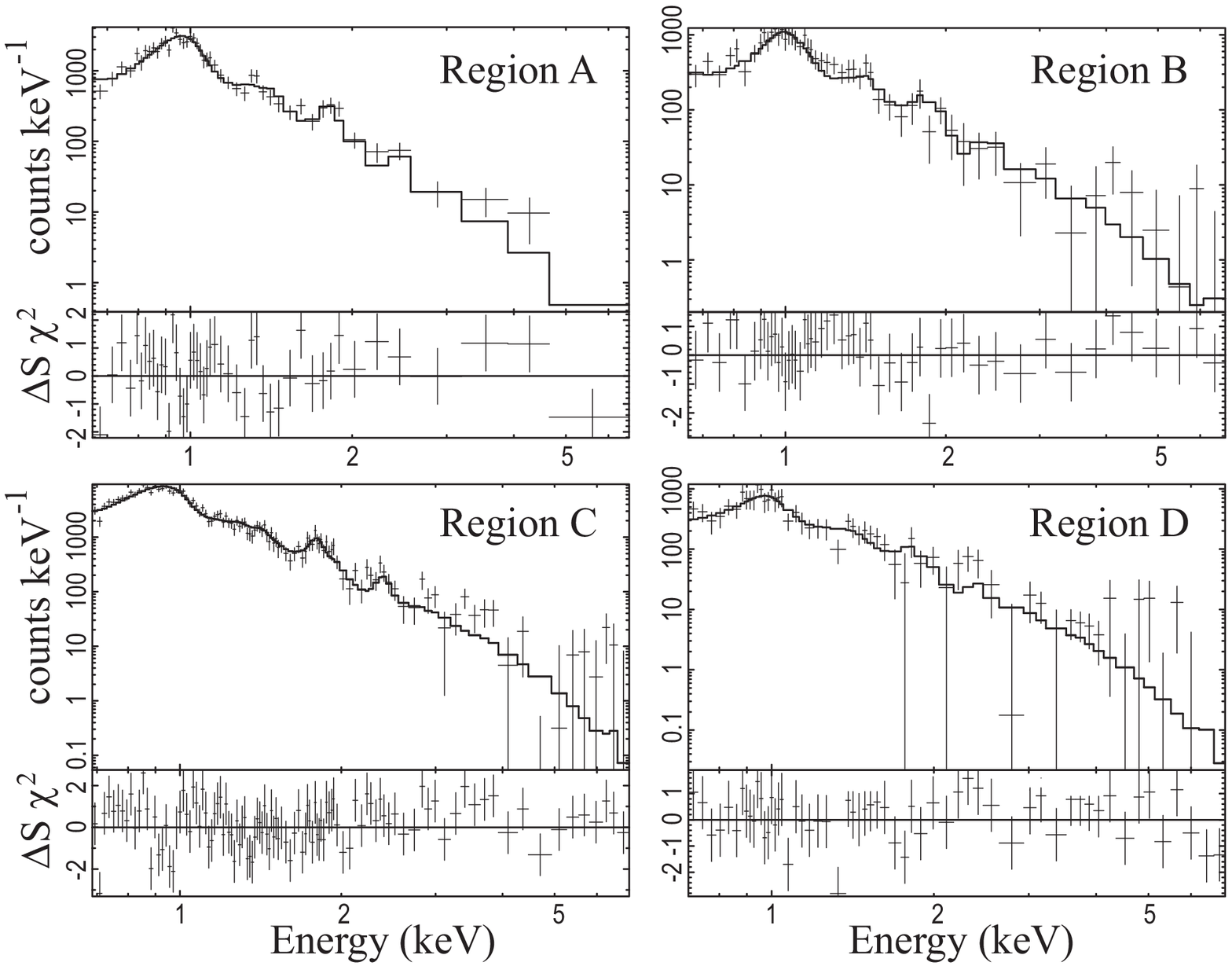}
\end{center}
\caption{Spectra extracted from the four regions defined in Fig. 2{\it
    c} and the best-fit models (see also Table 1).}
\end{figure}

In order to verify the reliability of our detection of the
high-abundance arc, we extract the spectra from regions A and B that
are defined on the arc, and regions C and D that are defined outside
the arc (Fig. 2{\it c}). The spectra are fitted with an absorbed APEC
model (Fig. 3), with the absorption fixed to the galactic value
$N_H=1.05\times10^{20}$ cm$^{-2}$ (Dickey \& Lockman, 1990), and the
redshift fixed to 0.0377 again. We summarize the best-fit results in
Table 1, which appear to be consistent with those implied in the
temperature and abundance maps (Fig. 2). By studying the
two-dimensional fit-statistic contours of gas temperature and metal
abundance at 68\% and 90\% confidence levels for regions A, B, C, and
D (Fig. 4), we are confident that the metal abundance in the arc
region is higher than those of the neighboring regions at a
significance of 90\%.  The existence of the high-abundance arc is
hence confirmed.

%%%%%%%%%%%%%%%%
%%Table 1%%%%%%%
%%%%%%%%%%%%%%%%
\begin{table}[!!!h]
   \begin{center}
     \caption{Best-fit spectral models for the four regions defined in
       Fig. 2{\it c}$^{\dag}$}
  \begin{tabular}{cccc}
    \hline\hline
    Region No. & Temperature (keV) & Abundance
    ($Z_\odot)$&$\chi^2/dof$\\
    \hline
    A&$1.05\pm0.03$&$0.83_{-0.24}^{+0.57}$&$49.4/45$\\
    B&$1.32_{-0.07}^{+0.09}$&$0.87_{-0.35}^{+0.85}$&$38.07/49$\\
    C&$0.92_{-0.03}^{+0.02}$&$0.40_{-0.08}^{+0.10}$&$121.02/106$\\
    D&$1.14_{-0.12}^{+0.14}$&$0.27_{-0.10}^{+0.13}$$^\ddag$&$64.98/56$\\
    \hline\hline
  \end{tabular}
\begin{minipage}{.86\textwidth}
  $^{\dag}$~We fit the spectra with an absorbed APEC model with the absorption and redshift fixed to the Galactic value ($N_H=1.05\times10^{20}$ cm$^{-2}$; Dickey \& Lockman 1990) and $z=0.0377$.\\
  $^{\ddag}$~The confidence level is 68\% for the abundance of region
  D.
  
\end{minipage}
\end{center}
\end{table}

\begin{figure}[h!!!]
\begin{center}
\includegraphics[width=12.5cm,angle=0]{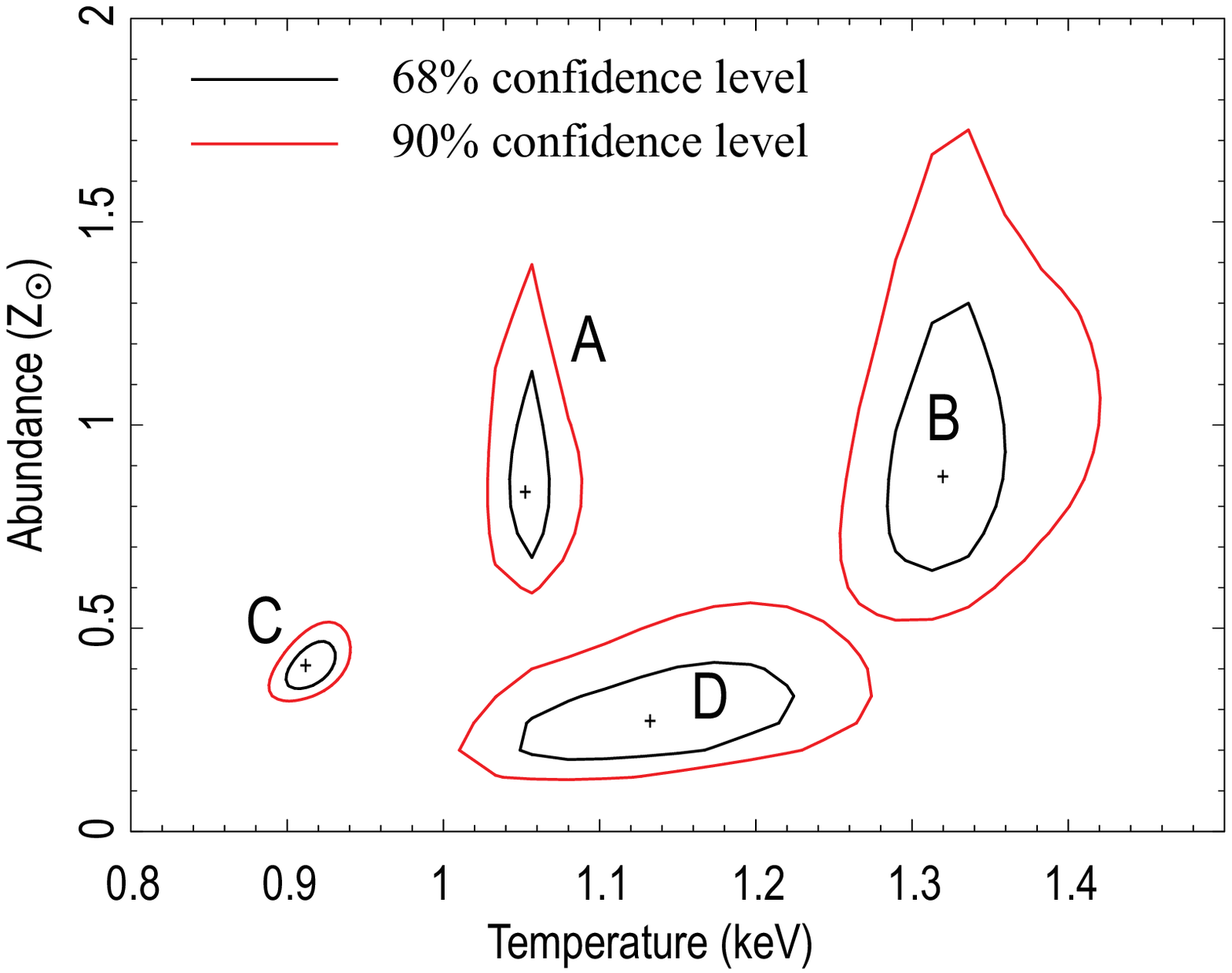}
\end{center}
\caption{ Two-dimensional fit-statistic contours of temperature and
  abundance at the 68\% and 90\% confidence levels for the four
  regions defined in Fig. 2{\it c}. }
\end{figure}

In order to compare the high-abundance structure with the spatial
distributions of X-ray emission and star lights, in Figure 2{\it c} we
overlay the X-ray surface brightness contours on the abundance map,
and mark the locations of the central galaxy PGC 046515 and member
galaxy PGC 046529 with two crosses. We find that the arm-like X-ray
feature (\S3) coincides with the high-abundance arc perfectly, and the
member galaxy PGC 046529 is located right at one end of the
high-abundance arc or the arm-like feature. These spatial correlations
indicate that both the arm-like X-ray feature and high-abundance arc
may be connected with PGC 046529.

\section{Discussion}
We detect a high-abundance arc in the RGH 80 compact group, which
shows tight spatial correlations with both the arm-like X-ray feature
and the member galaxy PGC 046529. A reasonable and natural way to
explain these correlations is to assume that the excess emission in
the arm-like X-ray feature is due to the excess iron contained in the
high-abundance arc, which originated from PGC 046529. But can this be
real? Firstly, we note that the gas temperature ($1.1-1.3$ keV) and
abundance ($\simeq0.7$ $Z_\odot$) of the high-abundance arc fall into
the temperature and abundance range of early-type galaxies. On the
other hand, considering that the metal abundance in the arc is about
$0.3$ $Z_\odot$ higher than its surrounding regions, and using the gas
density profile given in Xue et al. (2004), we estimate that the total
gas mass of the arm-like feature is $M_{\rm gas}\simeq5.4\pm
0.5\times10^{10}$ $M_\odot$, so that the excess iron mass therein is
$M_{\rm Fe}\simeq4.7_{-0.7}^{+6.5}\times10^6$ $M_\odot$. Assuming the
time dependent type Ia supernova rate in Dahlen et al. (2004) and the
wind mass loss rate in Ciotti et al. (1991), these amounts of gas and
iron are close to the lower limit of gas and iron mass of the
inter-stellar medium (ISM) of early-type galaxies, and can easily be
produced in PGC 046529 within only $\simeq 0.4$ Gyr, which strongly
supports the idea that the high-abundance gas was originally contained
in PGC 046529. Since there is no signatures of AGN activity (e.g.,
X-ray cavity and radio substructures) in PGC 046529, the possibility
of the asymmetric high-abundance arc being formed by AGN activity can
be excluded immediately. Hence the ram pressure stripping due to the
motion of PGC 046529 becomes the most hopeful mechanism.

By examining Figure 2, we also note that the gas temperature in the
high-abundance arc varies from $\simeq1.3$ keV in the north (region B)
to $\simeq1.1$ keV in the east (region A). Using the best-fit spectral
parameters from region A that are presented in Table 1 and Eq. (6) of
Gu et al. (2009), we estimate that the time needed for the low
temperature gas stripped out of PGC 046529 to achieve a thermal
equilibrium with the surrounding IGM via thermal conduction is $t_{\rm
  cond}\simeq 0.2$ Gyr. On the other hand, based on the velocity
dispersion ($450$ km s$^{-1}$; Ramella et al. 2002) of this group, we
roughly estimate that the time needed for PGC 046529 to travel a
distance equal to the separation between region B and region A is
$t_{\rm B\to A}\simeq0.15$ Gyr. Since $t_{\rm cond}\simeq t_{\rm B\to
  A}$ and PGC 046529 is located in region A, the spatial variation of
gas temperature from region B to region A strongly supports the idea
that the high abundance gas was stripped out of PGC 046529, and has
been conductively heated since then.

Compared with the two straight gas tails behind the infalling galaxy
ESO 137-001 in Abell 3627, as found by Sun et al. (2009), the spatial
distribution of the high-abundance gas in RGH 80 forms a curved shape
around PGC 046529, and possesses a width much wider than the gas tails
behind ESO 137-001. This could be ascribed to the low spatial
resolution ($\simeq0.8^\prime$ on the high-abundance arc) of our
abundance map. Besides, since the location of PGC 046529 from the
group center ($30.5h_{71}^{-1}$ kpc) is significantly less than the
Roche limit
\begin{equation}
  d\simeq2.44\left(\frac{M}{\frac{4}{3}\pi\rho_m}\right)^{1/3}\simeq182~h_{71}^{-1}~{\rm kpc},
\end{equation}
where $\rho_m\simeq5\times10^{-26}$ g cm$^{-3}$ is the density of the
gas stripped from PGC 046529, and $M\simeq 1.5\times10^{12}$ $M_\odot$
is the total gravitational mass contained inside the radius of PGC
046529 (Xue et al. 2004), it is possible that the high-abundance arc
has also been significantly broadened by the tidal force. Such an
off-center enrichment process may help explain the central metal
abundance dip observed in many galaxy clusters and groups (e.g., Abell
3266, Henriksen \& Tittley 2002; the NGC 1550 group, Sun et al. 2003).

Based on above analysis and discussion, we conclude that the
high-abundance gas around the member galaxy PGC 046529 in RGH 80 is
the remnant of the ISM of this galaxy, which was removed outwards by
the ram pressure due to the motion of PGC 046529. This novel case
shows that the ram pressure can serve as an efficient metal enrichment
mechanism in galaxy groups, just as in galaxy clusters.

\section{Summary}
With the high-quality \Chandra\ ACIS data, we detect a high-abundance
arc structure, in which the metal abundance is significantly higher
than the surrounding regions by $\simeq 0.3$ $Z_\odot$. This structure
shows tight spatial correlations with the member galaxy PGC 046529, as
well as with the arm-like feature identified on the X-ray image in
previous work of Randall et al. (2009).  Since no AGN activity is
found in PGC 046529, and the gas temperature, metallicity and mass
resemble those of the ISM of early-type galaxies, we conclude that the
high-abundance arc is the remnant of the ISM of PGC 046529, which was
efficiently stripped out of this galaxy by ram pressure.

%#########################
\section*{Acknowledgments}
%#########################
We thank the \Chandra\ team for making data available via the High
Energy Astrophysics Science Archive Research Center (HEASARC) at
http://heasarc.gsfc.nasa.gov. This work was supported by the National
Science Foundation of China (Grant No. 10673008, 10878001, and
10973010), the Ministry of Science and Technology of China (Grant
No. 2009CB824900/2009CB24904), and the Ministry of Education of China
(the NCET Program).

%\clearpage

\label{lastpage}

\end{document}